\documentclass[%
 reprint,
usenatbib,
 amsmath,amssymb,
 aps,
 nofootinbib, 
twocolumn]{aastex631}

\usepackage{sidecap}

\usepackage{rotating}
\usepackage{graphicx}
\usepackage{dcolumn}
\usepackage{tabularx}
\usepackage{bm}
\usepackage{hyperref}

\usepackage{booktabs, tabularx, makecell, multirow, xspace}
\usepackage{graphicx}	
\usepackage{amsmath}	
\usepackage{amssymb}	

\usepackage{physics}   
\usepackage{comment}   
\usepackage{slashed}

\newcommand{\Lsun}{{\,\rm L_\odot}}

\newcommand{\K}{\,{\rm K}}

\renewcommand{\arraystretch}{1.8}
\newcolumntype{C}[1]{>{\centering\let\newline\\\arraybackslash\hspace{0pt}}m{\#1}}

\def\aap{A\&A}

\def\apjl{ApJ}

\def\apjs{ApJS}

\newcommand{\rhocms}{\rho_\mathrm{core}}
\newcommand{\massms}{m_{\rm{MS}}}
\newcommand{\Tnugget}{T_{\rm{nugget}}}
\newcommand{\rhonugget}{\rho_{\rm{nugget}}}
\newcommand{\Tcnugget}{T_c^{\rm{nugget}}}
\newcommand{\rhocnugget}{\rho_c^{\rm{nugget}}}
\newcommand{\Mnugget}{M_{\rm{nugget}}}
\newcommand{\massnugget}{m_{\rm{nugget}}}
\newcommand{\Lphoto}{\mathcal{L}_{\rm{photo}}}

\newcommand{\Tphoto}{T_{\rm{photo}}}
\newcommand{\rphoto}{r_{\rm{photo}}}

\newcommand{\Gaia}{$Gaia$\ }

\newcommand{\DC}[1]{\textcolor{blue}{[DC: #1]}}

\newcommand{\sref}[1]{Section~\ref{#1}}

\newcommand{\fref}[1]{Figure~\ref{#1}}
\newcommand{\eref}[1]{Eqn.~\eqref{#1}}

\newcommand\blfootnote[1]{%
  \begingroup
  \renewcommand\thefootnote{}\footnote{#1}%
  \addtocounter{footnote}{-1}%
  \endgroup
}

\begin{document}


\title{Generalized Predictions for the Electromagnetic Signatures of Mirror Stars}

\author[0009-0007-6663-5921]{Franco Cabral*}
\affiliation{Department of Physics, University of Toronto, Toronto, Ontario M5S 1A7, Canada}

\author[0009-0006-1966-1492]{Stuart Williamson*}
\affiliation{Department of Astronomy and Astrophysics, University of Toronto, Toronto, Ontario M5S 1A7, Canada}

\author[0000-0003-0263-6195]{David Curtin}
\affiliation{Department of Physics, University of Toronto, Toronto, Ontario M5S 1A7, Canada}

\author[0000-0001-9732-2281]{Christopher D.~Matzner}
\affiliation{Department of Astronomy and Astrophysics, University of Toronto, Toronto, Ontario M5S 1A7, Canada}

\begin{abstract}
Mirror Stars are a generic prediction of dissipative dark matter models, including minimal atomic dark matter and twin baryons in the Mirror Twin Higgs. 
Mirror Stars capture regular atoms from the interstellar medium through highly suppressed kinetic mixing interactions between the regular and the dark photon. This results in the accumulation of a `` nugget'', which draws heat from the mirror star core and emits distinctive X-ray and optical signals.
In this work, we solve the stellar structure equations of optically thick nuggets across a wide range of the effective mirror star parameter space, and characterize their emission spectra using stellar atmosphere models. 
This complements an earlier analysis of lower-mass optically thin nuggets. 
We find that optically thick mirror star nuggets occupy distinct regions of the (stellar surface temperature, luminosity, surface gravity) space, and can be distinguished from regular stars in both HR diagrams and temperature-surface-gravity diagrams using astrometric and spectroscopic stellar catalogues.
Our detailed predictions, which are publicly available, now
give for the first time a general picture of mirror star signals in the optical and IR to enable  realistic mirror star searches using existing catalogues and new  telescope observations.
\end{abstract}

\section{Introduction}
\label{sec:intro}
\blfootnote{\emph{First authors are indicated with *.}}

Although the presence of dark matter (DM) is firmly established, its precise properties and possible interactions remain largely mysterious.  The standard picture is that of cold dark matter (CDM), which is generally treated as collisionless on astrophysical length scales and is often modeled as a single novel particle species, such as the Weakly Interacting Massive Particle (WIMP)~\citep{Battaglieri:2017aum}.  This framework successfully explains the observed large-scale distribution of matter in the universe~\citep{Planck:2018vyg}, and it reproduces the unseen mass component inferred from gravitational effects in galaxies and clusters. Despite these successes, and the appealing minimalism of the CDM hypothesis, there are many compelling reasons to consider more elaborate possibilities beyond this baseline scenario.

Observational tensions between CDM-based simulations and data on dwarf galaxy scales~\citep{Bullock:2017xww,Governato:2009bg,Garrison_Kimmel_2019,Relatores_2019} have long hinted that dark matter might exhibit nontrivial self-interactions, although the evidence is evolving rapidly with the discovery of many ultra-faint dwarf galaxy satellites of the Milky Way \citep{Homma2024,Kahenisa24}.
On theoretical grounds, it is quite natural to expect that DM could belong to a richer dark sector containing multiple particles and forces, a framework that can naturally address such small-scale anomalies~\citep{Tulin:2013teo}.  
From a bottom-up perspective, this expectation is reinforced by the fact that the Standard Model (SM) itself is far from minimal in its particle content and interactions.  
A useful way to explore this idea is through simplified models such as Atomic Dark Matter (aDM)~\citep{Kaplan:2009de}, which posit a hidden analogue of electromagnetism with an unobserved, massless dark photon, under which an asymmetric relic abundance of two DM species, a heavier “dark proton” and a lighter “dark electron”, carry opposite charges.

Crucially, a wide range of beyond-the-Standard-Model (BSM) scenarios that address persistent puzzles in fundamental physics predict exactly such dark sectors.  
Prominent examples include Neutral Naturalness frameworks such as the Mirror Twin Higgs \citep[MTH:][]{Chacko:2005pe, Chacko:2016hvu, Craig:2016lyx} and NNaturalness \citep{Arkani-Hamed:2016rle}, which provide alternative solutions to the electroweak hierarchy problem compared to more traditional approaches like supersymmetry~\citep{Martin:1997ns}.  
Because they do not rely on the same mechanisms as canonical models, these theories remain consistent with the lack of new physics signals at the Large Hadron Collider; see the reviews in \citet{Batell:2022pzc} and \citet{Craig:2023lyt}.  
In both MTH and NNaturalness, one finds at least one complex dark sector related to the SM by a discrete symmetry.  
This leads to the prediction of twin analogues of protons, neutrons, and electrons, interacting through mirror copies of SM gauge forces, thereby realizing aDM-like dynamics but extended with nuclear interactions.  
Moreover, the discrete symmetry strongly suggests that aDM (or a similar construction) should plausibly account for only a \emph{subset} of the total DM density: since the mechanism that produces ordinary baryons in the early universe, responsible for roughly one-sixth of the cosmic matter budget, would generically produce an analogous dark baryon component as well (see, for example,~\cite{Farina:2015uea, Earl:2019wjw, Alonso-Alvarez:2023bat}).  
Such sub-dominant aDM contributions remain consistent with existing limits on DM self-interactions~\citep{Randall:2008ppe}, even if their phenomenology differs substantially from CDM.

Beyond being self-interacting, aDM can also assemble into bound atomic states and is inherently \emph{dissipative}: it is able to radiate dark photons, thereby cooling and undergoing gravitational collapse to form structure.  
As a result, aDM can leave characteristic signatures in early-universe cosmology~\citep{Cyr-Racine:2013fsa, Bansal:2022qbi, Bansal:2021dfh, Zu:2023rmc, Barron:2026nks, Adams:2026tks}, large-scale structure~\citep{Bose:2018juc, Barron:2025dys}, and galactic dynamics~\citep{Fan:2013yva, Ghalsasi:2017jna, Roy:2023zar, Gemmell:2023trd}.\footnote{For additional discussions, see~\cite{Fan:2013tia,McCullough:2013jma,Randall:2014kta,Schutz:2017tfp, Buch:2018qdr,Ryan:2021dis, Gurian:2021qhk, Ryan:2021tgw, Foot:2013lxa,  Foot:2014uba, Foot:2015mqa,Chashchina:2016wle, Foot:2017dgx, Foot:2018dhy, Foot:2016wvj, Foot:2013vna}.}
Among the most striking consequences of this framework is the possible formation of \emph{Mirror Stars}.

Mirror Stars are, by direct analogy with ordinary stars, gravitationally bound compact objects composed of aDM, which radiate away their internal energy through the emission of invisible dark photons. Depending on additional details of the aDM model, these mirror stars could be fueled by internal conversions analogous to nuclear fusion (if there is dark nuclear physics), or they could simply cool on their Kelvin-Helmholtz timescale. 
These mirror stars will accrete ordinary matter from the interstellar medium through highly suppressed scattering processes, mediated by a tiny kinetic mixing which is expected to be present between the dark and SM photon~\citep{Gherghetta:2019coi}. 
The captured SM material quickly sinks to the stellar core, in a manner reminiscent of the proposed capture of DM in the Sun~\citep{1985ApJ...296..679P,1987ApJ...321..571G}.  
It was shown~\citep{Curtin:2019ngc,Curtin:2019lhm} that this ordinary matter is heated in the core 
and emits distinctive optical and X-ray radiation.  
Even when this emission is expected to be faint, the resulting luminous \emph{nuggets} could still be observed within distances of order $\mathcal{O}(100$--$1000\,\mathrm{pc})$, opening a novel  observational window onto atomic dark matter through astrophysical surveys.
While the local mirror star abundance is highly model-dependent and exceedingly difficult to predict for a given microphysical aDM Lagrangian, benchmark aDM simulations~\cite{Roy:2023zar} demonstrate that local mirror star mass densities of order $\mathcal{O}(M_\odot/\mathrm{pc}^3)$ can easily be realized even for percent-level DM mass fractions of aDM. Directly  observing mirror stars is clearly a tantalizing discovery opportunity.

Searching for mirror stars in public \Gaia~\citep{Howe:2021neq} and other catalogue data requires detailed predictions for the spectrum of the electromagnetic radiation emitted by the captured nuggets. 
Two distinct cases must be studied: low-mass optically thin nuggets, and higher-mass optically thick nuggets. 
In a previous publication~\citep{Armstrong:2023cis}, the former case was analyzed in detail by adapting the nebular synthesis code \texttt{\texttt{Cloudy}} \citep{ferland1998cloudy} to the case of optically thin gas clouds in the gravitational well of a mirror star. 
That analysis also established that the properties of the captured nugget only depend on three effective parameters: the nugget mass $M_\mathrm{nugget}$, the mirror star central core density $\rho_\mathrm{core}$, and an effective heating rate $\xi$. 
This parameterization separates a detailed treatment of the nuggets from the potentially unknown mirror star microphysics, and allows the parameter space of possible optical mirror star signals to be explored exhaustively in detail. The predicted spectra and magnitudes for optically thin nuggets are publicly available.\footnote{\url{https://github.com/davidrcurtin/mirror_star_emissions} \label{foot.github}} There are broad regions of this parameter space where mirror stars are observable.

In this paper, we study the second case of higher-mass optically thick nuggets captured by mirror stars. 
We solve for the nugget's hydrostatic structure, determine its spectrum of surface emissions, and analyse how these signals compare to the  emissions from regular SM stars. 
We again make the results publicly available.\textsuperscript{\ref{foot.github}} Together with the results for optically thin nuggets, this completes our picture of optical/IR mirror star signals and enables a wide range of new searches for mirror stars. 

This paper is structured as follows. \sref{sec:review} briefly reviews the physics of mirror stars, their capture of SM matter, the resulting electromagnetic emissions and the effective parameterization of the mirror star signal.
In \sref{s.modelingnugget} we outline how we solve for the structure and surface temperature of optically thick nuggets. This is combined with stellar atmosphere models in \sref{s.emsignaturesmirrorstars} to predict the detailed mirror star emission spectra from nugget surface emission, and the signal is compared to the emissions of regular stars. We conclude with a brief discussion of our results and their implications in \sref{s.conclusion}.

\vspace{5mm}

\section{Mirror Star Review \label{sec:review}}

We now provide a short review of the relevant physics of mirror stars. For a more detailed discussion, the reader is directed to Section 2 of the  \cite{Armstrong:2023cis} companion paper on optically thin nuggets, as well as the various original references.

Mirror star formation is a direct prediction of almost any aDM scenario~\cite{Curtin:2019ngc, Curtin:2019lhm} (see also~\cite{Foot:1999hm, Foot:2000vy, Foot:2004pa, Foot:2014mia}), and proceeds through cooling and collapse in direct analogy to regular star formation. 
The mirror star mass function is in principle a function of the underlying aDM parameters, though it can also be greatly modified by additional features of the microphysics not included in the aDM parameterization, like the possibility of dark nuclear physics. Similarly, the lifetime can vary widely, and be set by either the Kelvin-Helmholtz timescale or the details of dark nuclear fusion. 
Finally, our results apply to 'main sequence' mirror stars as well as degenerate objects representing end points of mirror star evolution, like mirror white dwarfs~\citep{Ryan:2022hku} or mirror neutron stars~\citep{Hippert:2021fch, Hippert:2022snq}. 
Fortunately, our study of mirror star electromagnetic signatures does not depend on those details.\footnote{Note however that connecting our results to complementary mirror star probes~\citep{Winch:2020cju, Perkins:2025hfr, Hippert:2021fch, Hippert:2022snq, Pollack:2014rja, Shandera:2018xkn, Singh:2020wiq, Gurian:2022nbx, Fernandez:2022zmc} from gravitational wave observations and microlensing surveys will require a more detailed understanding of mirror star characteristics.} We are also agnostic as to the mirror star spatial distribution in our galaxy, though new simulation studies~\citep{Roy:2023zar, Roy:2024bcu} may begin to shed light on the issue.

If aDM and hence mirror stars exist, it is highly likely that the dark photon and the visible SM photon share a tiny \emph{kinetic mixing} interaction $\mathcal{L} \supset \frac{1}{2}\epsilon F_{\mu \nu} F_D^{\mu \nu}$, where $F, F_D$ are the field strength of the SM and dark photon respectively.
Formally $\epsilon$ can be of any size, but there is a well-motivated range. Dynamical contributions to the kinetic mixing are generated by any interactions between the dark and visible sector, including quantum gravity, which motivates $\epsilon \gtrsim 10^{-14} - 10^{-13}$~\citep{Gherghetta:2019coi}, while cosmological constraints generally require $\epsilon \lesssim 10^{-10}$~
\citep{Vogel:2013raa, Iles:2024zka}, which is compatible with predictions from, for example, certain UV completions of Minimal Twin Higgs models~\citep{Koren:2019iuv}.
Mixings in this range are cosmologically and astrophysically unconstrained, but lead to capture of SM matter from the interstellar medium (ISM) in mirror stars~\citep{Curtin:2019ngc, Curtin:2019lhm}.\footnote{Other effects include direct detection signals~\citep{Chacko:2021vin, SENSEI:2020dpa, SuperCDMS:2022kse} and stellar cooling~\citep{Curtin:2020tkm, Fung:2023euv}, which can provide complementary constraints.} The captured SM gas sinks to the center of the mirror star and forms a `nugget' with masses at the asteroid scale or above, depending on the properties and location of the mirror star. 
The nugget is heated up through kinetic-mixing-mediated interactions with the mirror star core and radiates in SM photons. Determining the properties of the nugget and hence the luminosity and spectral characteristics of the resulting
electromagnetic mirror star signal is the object of our analysis.

\cite{Armstrong:2023cis} proposed a useful effective parameterization which completely determines the properties of the nugget and its thermal emissions without direct reference to the aDM microphysics or the detailed properties of mirror stars. The three parameters are the \emph{nugget mass} $M_\mathrm{nugget}$, the \emph{mirror star central core density} $\rho_\mathrm{core}$, and the \emph{heating rate} $\xi$.\footnote{Note that in addition to thermal emissions, the nugget also radiates in X-rays~\citep{Curtin:2019ngc, Curtin:2019lhm}, which result from  elastic Compton conversion of dark photons in the mirror star core directly to visible photons by scattering off SM electrons. 
This is a fascinating smoking-gun signal of mirror stars in its own right, with the X-ray frequency set directly by the mirror star core temperature $T_\mathrm{core}$.
Extending our effective parameterization to also determine this X-ray signal would require adding $T_\mathrm{core}$ as a fourth parameter. We leave this to future work and focus on characterizing the nugget's optical/thermal emissions.}

Under the assumption that the nugget is much smaller than the mirror star, the mirror star density (which dominates the gravitational potential) can be approximated to be constant at its central value $\rho_\mathrm{core}$ on the scale of the nugget. 
Assuming radial symmetry, the total heating rate into the nugget as a function of distance from the center is then
\begin{equation}
    \label{f.totalheatingrate}
    \frac{d \mathcal{L}}{dV}(r) = \frac{X \rho_\mathrm{nugget}(r)}{m_H} 
    \left( \frac{\rho_{{\rm core}}}{\rho_{{\rm core}, \odot}}\right) \ 
     \xi \ 
    (\mathrm{J/s}) \textit{} .
\end{equation}
The first term is just the hydrogen number density in the nugget, with $X$ being the (assumed uniform) hydrogen mass fraction,  $\rho_\mathrm{nugget}(r)$ the nugget density, and $m_H$ the mass of a hydrogen atom. The second term is the mirror star core density normalized to the Sun's central density of $160
~\mathrm{g}/\mathrm{cm}^3$.
The heating rate $\xi$ is dimensionless (hence the explicitly included units $(J/\rm s)$ above); normalized to the hydrogen number density for consistency with~\cite{Armstrong:2023cis}; and relates to the dark photon kinetic mixing $\xi \propto \epsilon^2$, with the constant of proportionality  being within an order of magnitude of unity for roughly SM-like aDM and mirror star parameters.
It is therefore clear that the total heating rate into the entire nugget is $\mathcal{L} \propto M_\mathrm{nugget} \rho_\mathrm{core} \xi$, which in the steady state has to be its total luminosity as well. 

This dictates a clear strategy for computing the mirror star properties and emissions for given values of $(M_\mathrm{nugget}, \rho_\mathrm{core}, \xi)$. The structure of the nugget is computed from hydrostatic and energetic equilibrium; the total luminosity is uniquely set by the product of the three parameters; and the thermal emission spectrum is computed alongside the hydrostatic solution depending on whether the nugget is optically thin, radiating via volume emission, or optically thick, radiating via surface black-body emission analogous to tiny regular stars. The former case was studied by \cite{Armstrong:2023cis}, while this paper focuses on the latter. More details on how to solve for the nugget properties will be given in the following sections.

While $\rho_\mathrm{core}$ depends in a complicated way on aDM microphysical parameters and the mass of a particular mirror star, we can easily gain some intuition into what kinds of nugget masses might be reasonable to expect.
For aDM microphysical parameters and mirror star properties within several orders of magnitude of their SM counterparts, and kinetic mixings in our range of interest $\epsilon \sim 10^{-14} - 10^{-10}$,
\cite{Curtin:2019ngc} found that the dominant capture mechanism very quickly becomes geometrical self-capture of ISM atoms colliding with the already captured SM nugget. 
We can therefore arrive at a fairly robust approximation of the capture rate as a function of the total mirror star mass $M_\mathrm{MS}$ using only some rough estimates of the nugget size.

The rough scale height of the nugget is 
\begin{equation}
    \label{e.hnugget}
    h_{{\rm nugget}} \sim (10^4~\mathrm{km}) \left[
    \left(\frac{T_\mathrm{nugget}}{10^5 \mathrm{K}}\right)\left( \frac{m_H}{\bar m}\right) \left(\frac{\rho_{\mathrm{core}, \odot}}{\rho_\mathrm{core}}\right)
    \right]^{1/2}
    \ .
\end{equation}
For the optically thin nuggets studied by \cite{Armstrong:2023cis}, the nuggets are approximately isothermal with $T_\mathrm{nugget} \sim 10^4~\mathrm{K}$, which is set by SM ionization energies. For the optically thick nuggets we study, $T_\mathrm{nugget}$ should be taken to be the internal temperature, which we find to be $\sim 10^{4} - 10^{6}~\mathrm{K}$ for nuggets that are fainter than but within a few orders of magnitude of solar luminosity. 
This small size compared to the extent of typical stars lends credence to our simplifying assumption of only considering the central mirror star core density. 

We can now estimate the mirror star's geometrical capture rate of ISM atoms. Updating the estimate of~\cite{Armstrong:2023cis} to include the effect of gravitational focusing~\citep{Petraki:2013wwa}, the capture rate is $dN/dt = \sqrt{3/2}\ n_i^\mathrm{ISM} \pi r_\mathrm{nugget}^2 \  \bar v_{esc}^2/u $, where $\bar v_{esc}$ is the escape velocity averaged over the nugget, $u$ is the velocity of incoming ISM atoms, and $r_\mathrm{nugget}$ is the radius of the nugget which is opaque to incoming atoms.
If the nugget is much smaller than the mirror star, the escape velocity at the nugget can be taken to be the escape velocity from the star center. 
Making the simplifying assumption that the mirror star has an approximately Gaussian density profile, we can therefore estimate
\begin{equation}
    \bar v_{esc} \sim (1500~\mathrm{km}/\mathrm{s}) \left(\frac{M_\mathrm{MS}}{M_\odot}\right)^{1/3} \left(\frac{\rho_\mathrm{core}}{\rho_{\mathrm{core},\odot}}\right)^{1/6} 
\end{equation}
for a total mirror star mass $M_\mathrm{MS}$.
To compute the total capture rate, we assume an interstellar medium density of $n_H^\mathrm{ISM} = 1~\mathrm{cm}^{-3}$ and take $u$ to be set by the one-dimensional velocity dispersion of  $\sim$ 30\,km/s characteristic of our local stellar environment~\citep[][Table 1.2]{binney2011galactic}. 
Taking $r_\mathrm{nugget}$ to be roughly constant throughout the nugget's evolution, since it is dominantly set by the external gravitational potential, gives the following expected nugget mass for a mirror star age $\tau_\mathrm{MS}$:
\begin{eqnarray}
 \nonumber   M_\mathrm{nugget} &\sim& \left(10^{22}~\mathrm{g}\right) 
    \left( \frac{\tau_\mathrm{MS}}{10^8~\mathrm{years}}\right)
    \left(\frac{r_\mathrm{nugget}}{10^5 ~\mathrm{km}}\right)^{-2} \times
    \\
    &&
    \label{e.Mnuggetcapturetime}
    \left( \frac{M_\mathrm{MS}}{M_\odot} \right)^{2/3}
    \left(\frac{\rho_\mathrm{core}}{\rho_{\mathrm{core},\odot}}\right)^{1/3} \ .
\end{eqnarray}
In our analysis, we consider benchmark mirror star core densities $\rho_\mathrm{core}$ in the range $10^{-5} - 10^{5}$ $\times \rho_{\mathrm{core}, \odot}$. 
As we discuss in \sref{s.nuggetproperties}, all the optically thick nugget solutions we find are plausible targets for observation assuming capture time scales of less than 10 billion years.

\section{Modeling a Captured Optically Thick Nugget}
\label{s.modelingnugget}

We use the standard stellar structure equations of hydrostatic equilibrium, energetic equilibrium, and mass conservation to solve for the nugget density and temperature profiles $\rho(r), T(r)$ for given mirror star properties: core density $\rho_\mathrm{core}$, heating rate $\xi$ and nugget central conditions $\rho(0), T(0)$. 
We assume that the nugget composition is that of the ISM. Specifically, we assume hydrogen, helium and metal mass fractions of $(X,Y,Z) = (0.7116, 0.2789, 0.0095)$, which are also the default \texttt{Cloudy} abundances and the same composition used in the study of optically thin nuggets by \cite{Armstrong:2023cis}.
As we are studying optically thick nuggets, we assume that the radiation diffusion equation holds, and use the radial optical depth as a check of self-consistency. We also assume that convective regions have nearly constant entropy, as is usually the case in stellar convective regions, allowing the adiabatic temperature gradient to be used.

Given mirror star parameters ($\rho_\mathrm{core}$ and $\xi$), we compute a family of hypothetical nugget solutions with different central conditions $\rho(0)$ and $ T(0)$, within which we must identify the locus of solutions that radiate as much power as they receive from the mirror star. 
For each nugget solution, the nugget mass $M_\mathrm{nugget}$ and luminosity $\Lphoto$ is calculated. The latter would formally require matching onto a stellar atmosphere model. For simplicity, we instead identify the photosphere as the location where a model's optical depth to the surface equals 2/3, and identify the blackbody luminosity at that radius.
In equilibrium, the photospheric luminosity must match the total power transferred from the mirror star to the nugget $\cal L \propto M_\mathrm{nugget} \xi$, see \eref{f.totalheatingrate}. Requiring $\cal L = \Lphoto$ uniquely determines the correct $\rho(0)$ initial condition for a given $T(0)$ initial condition and hence uniquely determines the solution for a given $\{\rho_\mathrm{core}, \xi,  M_\mathrm{nugget}\}$. We can then scan over this parameter space to map out the properties of mirror star nuggets in  generality.  
The custom Python code we use to find optically thick nugget solutions is publicly available on GitHub.\textsuperscript{\ref{foot.github}}

\subsection{Nugget Opacity}

The relevant opacities are Rosseland mean values, indicated by a subscript $R$.  At temperatures $T>4000K$, we use analytical approximations of $\bar \kappa_R$ for opacity sources: electron scattering, H$^-$, bound-free, and free-free. We combine opacities such that H$^-$-dominated opacity transitions to the others in a realistic way. We also apply the electron scattering opacity $\kappa_{es}$ only to the ionized fraction of hydrogen atoms $x_{H^+}$, which we determined from the Saha equation: 
\begin{equation}
    \label{e.Opacity}
    \bar\kappa_R = x_{H^+}\cdot\kappa_{es} + \min\{\bar\kappa_R^{\rm H^-}, \bar\kappa_R^{\rm bf}+ \bar\kappa_R^{\rm ff}\}
\end{equation}

Defining $\rho_{\rm{cgs}}={\rho}/{1\,\rm g\,cm^{-3}}$, the expressions for these opacity sources are:
\begin{align*}
    \kappa_{\mathrm{es}} &= 0.2(1+X) \ \rm cm^2/g\\
    \bar\kappa^{\rm H^-}_R &= 10\rho_{\rm{cgs}}^{1/2}\left(\frac{T}{3250K}\right)^9\ \rm \frac{cm^2}{g} \\
    %
    \bar\kappa_R^{\rm bf} &= \frac{Z}{0.02}\left(\frac{1 + X}{1.7}\right)\,\rho_{\rm{cgs}}\,\left(\frac{7.91\cdot10^6 K}{T}\right)^{7/2} \ \rm \frac{cm^2}{g} \\
    \bar\kappa_R^{\rm ff} &=(X + Y)(1 + X)\,\rho_{\rm{cgs}}\,\left(\frac{2.87\cdot10^6 K}{T}\right)^{7/2}\ \rm \frac{cm^2}{g}
\end{align*}
\citep[e.g.,][]{HansenKawalerTrimble04}.
For temperatures $75~\mathrm{K} < T < 4000~\mathrm{K}$, we use a table created by \cite{Freedman:2014} for use in giant planet and ultracool dwarf atmospheres. We use their solar metallicity $[M/H] = 0$ table.

\subsection{Nugget Internal Structure}

With $\rhocms,\xi$ fixed, the nuggets are entirely described by the radial profiles $\{\rho(r), T(r)\}$. These can be transformed into pressure profiles $\{P(r), P_{\rm rad}(r)\}$ using $P = P_{\rm gas} + P_{\rm rad}$. We assume gas pressure is ideal, so $P_{\rm gas} = \rho k_b T/\mu,$ and $P_{\rm rad} =aT^4/3$. (Here $a$ is the radiation constant, related to the Stefan-Boltzmann constant by $a={4\sigma}/{c}$.)

Since we assume that the mirror star is much larger than the nugget, the $\rhocms$ density is a constant. So the mirror star mass contained in a shell of radius $r$ is $\massms(r) = \frac{4}{3}\pi r^3 \rhocms$. The local gravitational acceleration is then $g(r) = G(\massnugget(r) + \massms(r))/r^2$ with $G$ the gravitational constant. Then we have the first stellar structure equation from hydrostatic equilibrium:
\begin{equation}
\label{e.Hydrostatic}
    \frac{dP}{dr} = -g\rho
\end{equation}

The radiation pressure equation depends on whether convection or radiation transports the energy. The convection criterion we use is Schwarzschild's: $\nabla_{\rm rad} > \nabla_{\rm ad}$ ~\citep{Kippenhahn_Weigert_Weiss_2012}, where these quantities are defined with $\beta = \frac{P_{\rm gas}}{P}$, so $1-\beta = \frac{P_{\rm rad}}{P}$: 
\begin{align}
    \nabla_{\rm rad} =\left(\frac{d\ln T}{d\ln P}\right)_{\rm rad} &= \frac{3}{16\pi cGa}\frac{\bar\kappa_RP\mathcal{L}}{(\massms+\massnugget)T^4} \\
    \nabla_{\rm ad} = \left(\frac{d\ln T}{d\ln P}\right)_{\rm ad} &= \frac{1+(1-\beta)(4+\beta)/\beta^2}{2.5+4(1-\beta)(4+\beta)/\beta^2}
\end{align}
\citep[see][]{kippenhahn1990stellar}.
Then the piecewise radiation pressure gradient is:
\begin{equation}
    \label{e.PradGrad}
    \frac{dP_{\rm rad}}{dr} = \begin{cases}
        -4g\rho\frac{P_{\rm rad}}{P} \nabla_{\rm ad}&  (\nabla_{\rm rad} > \nabla_{\rm ad})\\
        -\bar\kappa_R\rho\frac{\mathcal{L}}{4\pi r^2c} & (\nabla_{\rm rad} \leq \nabla_{\rm ad})
    \end{cases}
\end{equation}
With the radiative case being the standard radiation diffusion equation and the convective case derived below. 

Because we assume in convective regions that the entropy is constant, the adiabatic gradient can be taken as the total temperature gradient: $\nabla_{\rm ad} = \frac{d\ln T}{d\ln P} = \frac{P}{T}\frac{dT}{dP}$. Rearranging, $\frac{dT}{dP} = \frac{T}{P}\nabla_{\rm ad}$. Using the chain rule and hydrostatic equilibrium, $\frac{dT}{dr} = \frac{dT}{dP}\frac{dP}{dr} = \frac{T}{P}\nabla_{\rm ad}(-g\rho)$. Then we differentiate the radiation pressure definition to get $\frac{dP_{\rm rad}}{dr} = \frac{4}{3}aT^3\frac{dT}{dr} = -4g\rho\frac{P_{\rm rad}}{P}\nabla_{\rm ad}$, as in \eref{e.PradGrad}.

With Eqns.\ \eqref{e.Hydrostatic}, \eqref{e.PradGrad}, along with mass conservation and initial conditions $\{\Tcnugget,\rhocnugget\}\iff\{P_c,P_{\rm c,rad}\}$, we integrate the problem with an adaptive-step-size first-order solver to generate $\{P(r), P_{\rm rad}(r), m(r)\}$.

\subsection{Photospheric Boundary Condition} \label{s.photosphere}

To find the correct line of self-consistent nuggets in the $\{\Tcnugget, \rhocnugget\}$-plane, we need to compare the heating luminosity $\mathcal{L}$ to the emitted luminosity $\Lphoto$. We model the total emission of the nugget in this stage as a black body emitting from the photosphere $\Lphoto = 4\pi \rphoto^2 \sigma \Tphoto^4$. The photospheric radius is the point at which the optical depth from infinity, $\tau_R(r) = \int_\infty^r\bar\kappa_R\rho\ ds$, reaches $2/3$:
\begin{equation}\label{e.photoCondition}
    \tau_R(\rphoto) = \frac{2}{3}, 
\end{equation}
 adopting the value that holds in the Eddington gray approximation \citep{kippenhahn1990stellar}. 
Then $\Tphoto = T(\rphoto)$.
We take $\rphoto$ to be the effective radius of the nugget, and take the total mass of the nugget to be $\Mnugget = \massnugget(\rphoto)$.

Note that our internal structure calculation uses a radiative structure near the photosphere, which assumes a high optical depth. This assumption breaks down at our photosphere condition, and so is not entirely consistent. Ideally, we would couple the atmosphere model used in section \ref{s.atmosphere} to the internal structure model, but this is unlikely to greatly change our calculated photosphere location and temperature at our desired level of precision. 

\begin{figure}
    \centering
    \vspace*{-6mm}
    \includegraphics[width=0.38\textwidth]{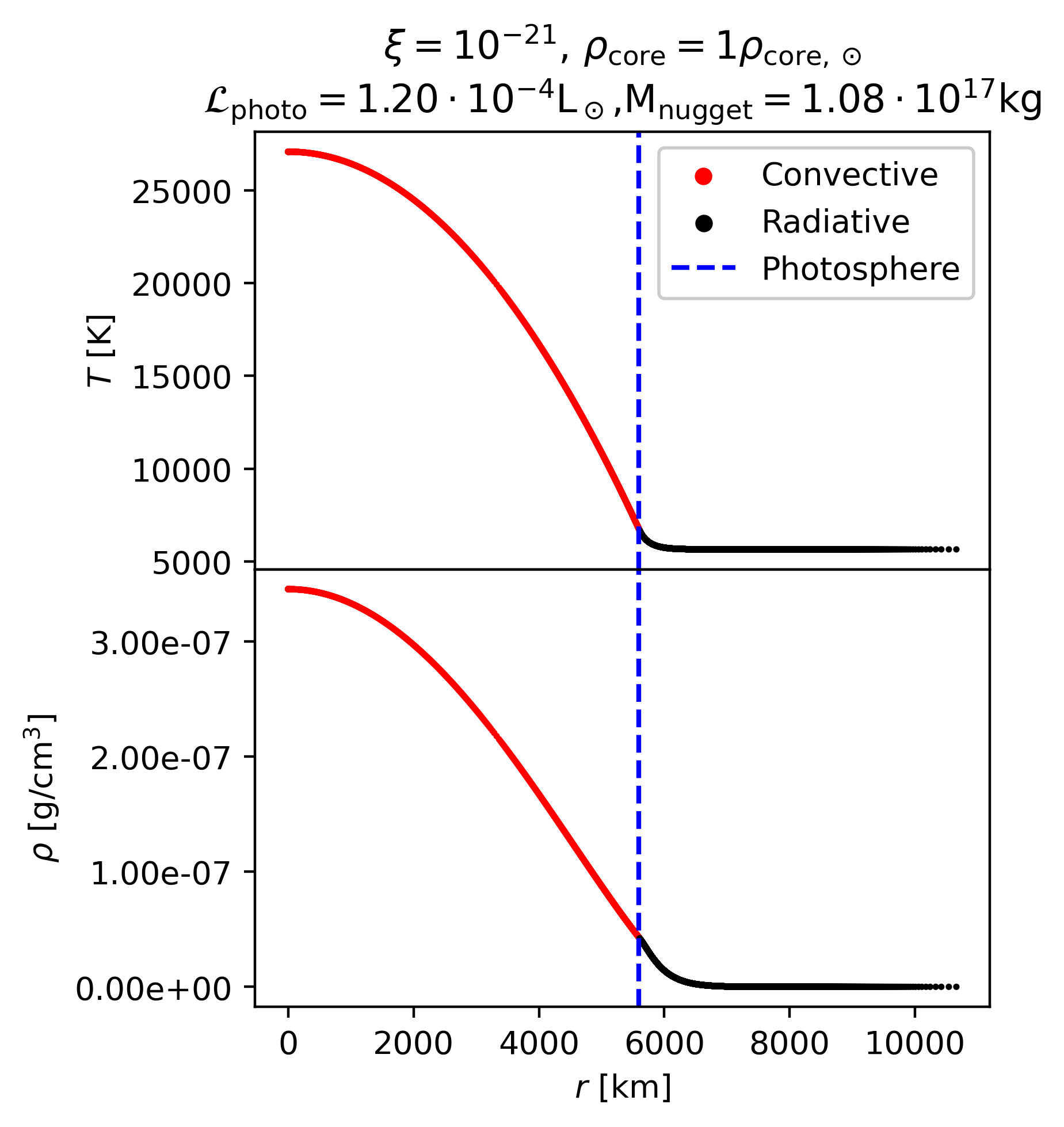}
    \includegraphics[width=0.38\textwidth]{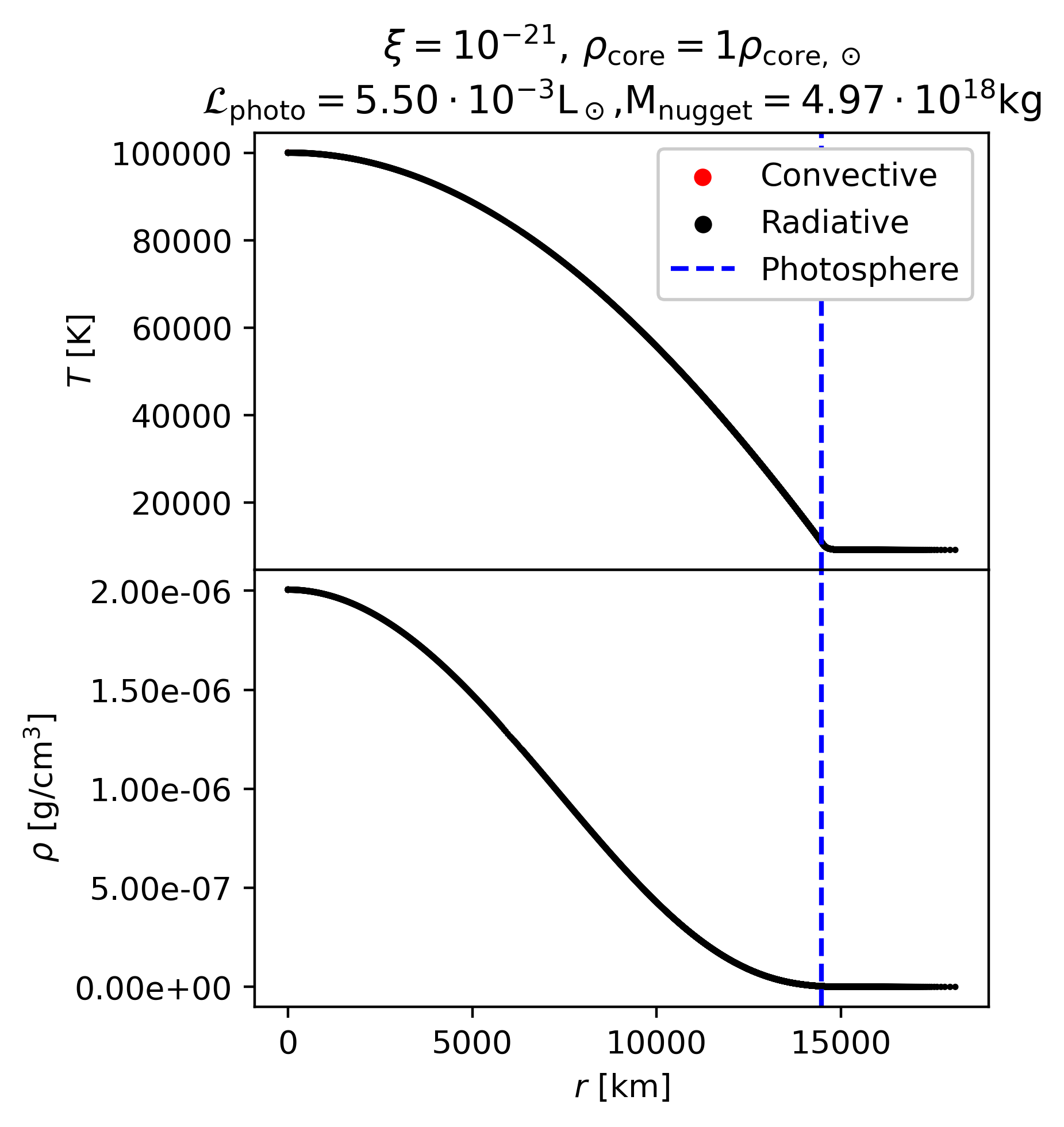}
    \caption{
        Profiles of $\Tnugget(r)$ and $\rhonugget(r)$ for two nuggets, one dominantly convective (top) and one dominantly radiative (bottom). Convective regions where $\nabla_{\rm{rad}}>\nabla_{\rm{ad}}$, see \eref{e.PradGrad}, are indicated in red. The dashed blue vertical line is the photospheric radius, where \eref{e.photoCondition} holds. $\xi, \rhocms, M_\mathrm{nugget}$ as well as the nugget luminosity $\Lphoto$ are listed at the top of each profile.}
    \label{f.profile1}
\end{figure}

\subsection{Properties of Optically Thick Nuggets}

\label{s.nuggetproperties}

Imposing the luminosity self-consistency condition $\cal L = \Lphoto$ uniquely selects the correct nugget density initial condition $\rho(0)$ for a given temperature initial condition $T(0)$. 
(In practice, we require our valid nugget solutions to satisfy $\cal L = \Lphoto$ to a precision of 0.1\%). 
Scanning over those three parameters $\{\rho_\mathrm{core}, \xi, M_\mathrm{nugget}\}$, we find optically thick nugget solutions satisfying $\tau_R(r=0) > 10$  across a wide volume of the mirror star parameter space.

\begin{figure*}
    \centering
    \vspace*{-6mm}
    \includegraphics[width=\textwidth]{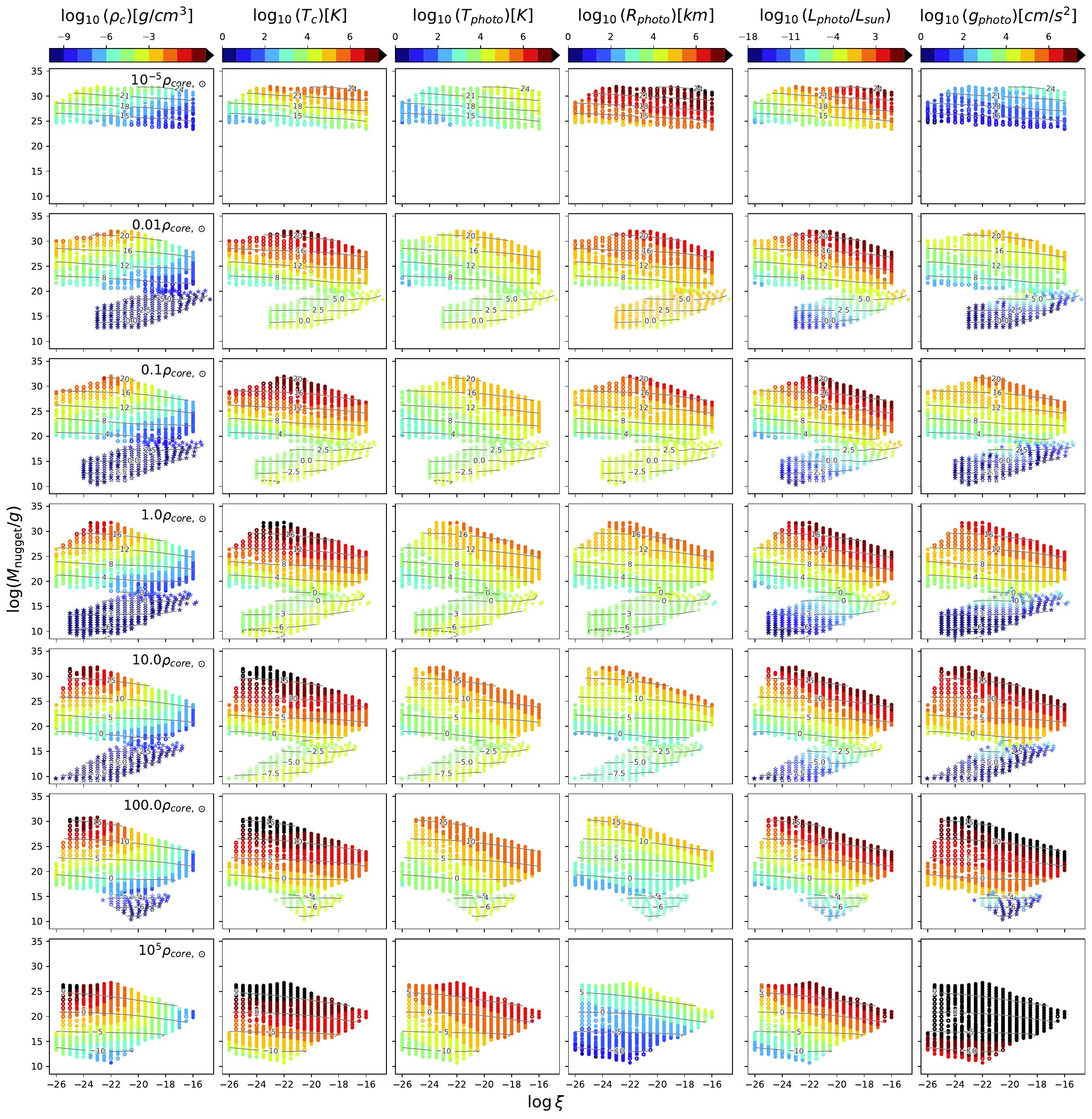}
    \caption{
        Nugget properties in the  in $(\xi,\Mnugget)$ plane, for 7 representative values of the mirror star core density $\rhocms$ relative to solar central density (different rows). 
        Left to right columns correspond to nugget central density, central temperature, photosphere temperature, photosphere radius, luminosity relative to the Sun, and surface gravity. 
        Optically thick nuggets are circular markers; optically thin nuggets from \cite{Armstrong:2023cis} are star markers.
        Hollow markers correspond to interpolated solutions in either regime.        
        Gray contours show $\log_{10}(\tau_\mathrm{MS,eff}/\mathrm{years})$, see \eref{e.tauMSeff}, indicating the required capture time scale of a solar mass mirror star to accumulate the nugget in the geometric capture regime, which is valid above the gray dashed contour. Note that this must be rescaled to the expected mirror star mass for the given central density. We do not include nuggets with a mass above $10^{32}~\mathrm{g}$, which is approximately the minimum mass for the nugget to become a hydrogen-burning red-dwarf even in the absence of the external mirror star gravitational potential, and we also exclude nuggets that make up more than 10\% of the mirror star mass within the nugget radius.  
    }
    \label{f.parameterspace}
\end{figure*}

We only consider nuggets with mass $M_\mathrm{nugget} < 10^{32}~\mathrm{g}$, since nuggets heavier than this will initiate fusion independently~\citep{Chabrier_2023} (even without the additional pressure from the mirror star gravitational well), which we did not account for in our modeling. 
We also discard nuggets with extremely high luminosities  $\mathcal{L} > 10^7\Lsun$, as stars this bright are not known to exist \citep{Ulmer:1997qj, Vink:2014lwa}. 
Finally, after imposing these cuts a handful of additional nuggets are removed by requiring that the mass of the nugget is subdominant compared to the enclosed mass of the mirror star: $M_\mathrm{nugget} < 0.1 m_{MS}(R_\mathrm{photo})$. This is not technically required, but makes it more likely that the nuggets we consider do not greatly affect mirror stellar evolution, which will simplify mapping a given dissipative dark matter model's predicted mirror star properties to our predicted signals.

In Fig \ref{f.profile1} (top) we show profiles of self-consistent nuggets for $\xi =10^{-21}, \rhocms=\rho_{\mathrm{core},\odot}$. \fref{f.profile1} shows a nearly wholly convective nugget, with the switch to a radiative outer shell just before the photosphere as mentioned in \sref{s.photosphere}. This nugget is low in luminosity, $\mathcal{L}\sim10^{-4}\Lsun$. 
Fig \ref{f.profile1} (bottom) by contrast shows an entirely radiative nugget, with a larger central temperature, central density, and luminosity $\mathcal{L}\sim5\cdot10^{-3}\Lsun$.

The central nugget density, central temperature, photosphere temperature, radius, luminosity and surface gravity of optically thick nuggets across the  $\{\rhocms, \xi, \Mnugget\}$ parameter space are shown in \fref{f.parameterspace} (circular markers). For comparison, we also show the optically thin solutions found in \cite{Armstrong:2023cis} (star markers). These nuggets slightly overlap with the optically thick nuggets, corresponding to regions where the optically thin assumption starts to fail, but mostly occupy their own region of the parameter space as expected.

The wide range of possible nugget properties makes it interesting to consider what regions of mirror star parameter space might be the most physically motivated. 
In \sref{sec:review}, we estimated the nugget mass from geometrical self-capture for some benchmark ISM parameters. In most of our parameter space of interest, both optically thin and thick nuggets are opaque enough to passing ISM atoms to be in the geometrical self-capture regime. The exception is some very low-mass optically thin nuggets. The lower boundary of the geometrical self-capture regime is indicated with a gray-dashed contour when applicable. 

In the geometrical self-capture regime we can estimate how long it would take for a mirror star to accumulate a nugget of a given mass. 
By inverting \eref{e.Mnuggetcapturetime}, we  can define
\begin{eqnarray}
\label{e.tauMSeff}
&& \tau_\mathrm{MS, eff} \equiv
\tau_\mathrm{MS} \left(\frac{M_\mathrm{MS}}{M_\odot}\right)^{2/3} 
\\
\nonumber
&& \sim
(10^8 \mathrm{years})
\left(\frac{M_\mathrm{nugget}}{10^{22} \mathrm{g}}\right) \left(\frac{r_\mathrm{nugget}}{10^5\mathrm{km}}\right)^2 \left(\frac{\rho_\mathrm{core}}{\rho_\mathrm{core, \odot}}\right)^{-1/3}
\end{eqnarray}
Gray contours of $\log_{10}(\tau_\mathrm{MS,eff}/\mathrm{years})$ are shown in \fref{f.parameterspace}.
For a given model of atomic dark matter, one could calculate the stellar properties of a mirror star, particularly $\rho_\mathrm{core}$ and $M_\mathrm{MS}$. In general, this can be very different from our SM stars, especially given that dark nuclear and electron masses can be orders of magnitude larger or smaller than in the SM, and should be the subject of a dedicated future study. However, we can anticipate, based on naive dimensional analysis and also the study of 'dark white dwarfs' by \cite{Ryan:2022hku}, that mirror star mass will generally vary inversely with the dark nucleon mass to an $\mathcal{O}(1)$ power, while central density will rise with a similar positive power. 

For mirror stars several orders of magnitude heavier than the Sun (which might occur if mirror nuclei are lighter than in the SM), we might therefore expect lower central densities. Consulting the $\tau_\mathrm{MS,eff}$ contours in the top two rows of  \fref{f.parameterspace}, and subtracting an $\mathcal{O}(1)$ number from the contour values to account for the mirror star mass, it would be plausible for nuggets with a 10 billion year accumulation timescale to still be optically thin, but it would also be possible for optically thick nuggets to form more quickly.
Conversely, if mirror stars are several orders of magnitude lighter than the Sun (which might occur if mirror nuclei are heavier), we consult the bottom two rows and add an $\mathcal{O}(1)$ number to the contours, leading us to expect most such mirror stars to have formed optically thick nuggets by today. 
Of course, whether those nuggets still shine also has to do with the longevity of mirror stars and their internal energy conversion mechanisms, if any, but broadly speaking, most of the optically thick nuggets in \fref{f.parameterspace} are motivated targets for mirror star searches.

It is worth briefly discussing the apparent gaps in the $(\xi, M_\mathrm{nugget})$-plane (for each $\rho_\mathrm{c}$) which are not populated by either optically thick or thin solutions. 
As discussed in \cite{Armstrong:2023cis}, regions below and to the right of the optically thin solutions, the heating rate is so high that solutions violate the non-relativistic assumptions of  the \texttt{Cloudy} code used in that work.  Regions below and to the left of the optically thin solutions are not populated by \texttt{Cloudy}, likely due to the steep dependence of the radiative cooling rate at low temperatures. 
There are also nugget mass ranges between the optically thin and our new optically thick solutions, at moderate to low heating rates, which are not populated by either method of obtaining nugget solutions. The reason that region is not populated by the analysis presented in this paper is a failure of the optical thickness criterion.  We therefore suspect that nuggets in this region represent a state between the completely optically thick and thin limits, which would require a dedicated analysis for this intermediate regime.

Finally, we point out that there are regions of parameter space within both the optically thin and thick regimes where no solutions are found for, as far as we can tell, purely numerical stability reasons which do not reflect a physical absence of stable nuggets with those parameters. Given the simple and smooth dependence of various observables on mirror star parameters, predictions can be obtained by interpolating across those regions. The corresponding interpolated nugget solutions are indicated in \fref{f.parameterspace} with hollow circle and star markers.

\section{Electromagnetic Signatures of Mirror Stars}

\label{s.emsignaturesmirrorstars}

We now describe the electromagnetic signatures of optically thick nuggets in Mirror Stars. This will establish a clear mirror star signal region in the HR diagram, but also establish purely spectral methods of distinguishing mirror stars from other astrophysical objects like white dwarfs and planetary nebulae. 

 Note that we ignore the possibility of chemical differentiation due to gravitational settling.  This is reasonable insofar as many of our models are convective up to the immediate vicinity of the photosphere.  However, because some models are fully radiative (see Figure  \ref{f.profile1}), the possibility of differentiation is worth revisiting.

\begin{figure}
    \centering
    \includegraphics[width=0.5 \textwidth]{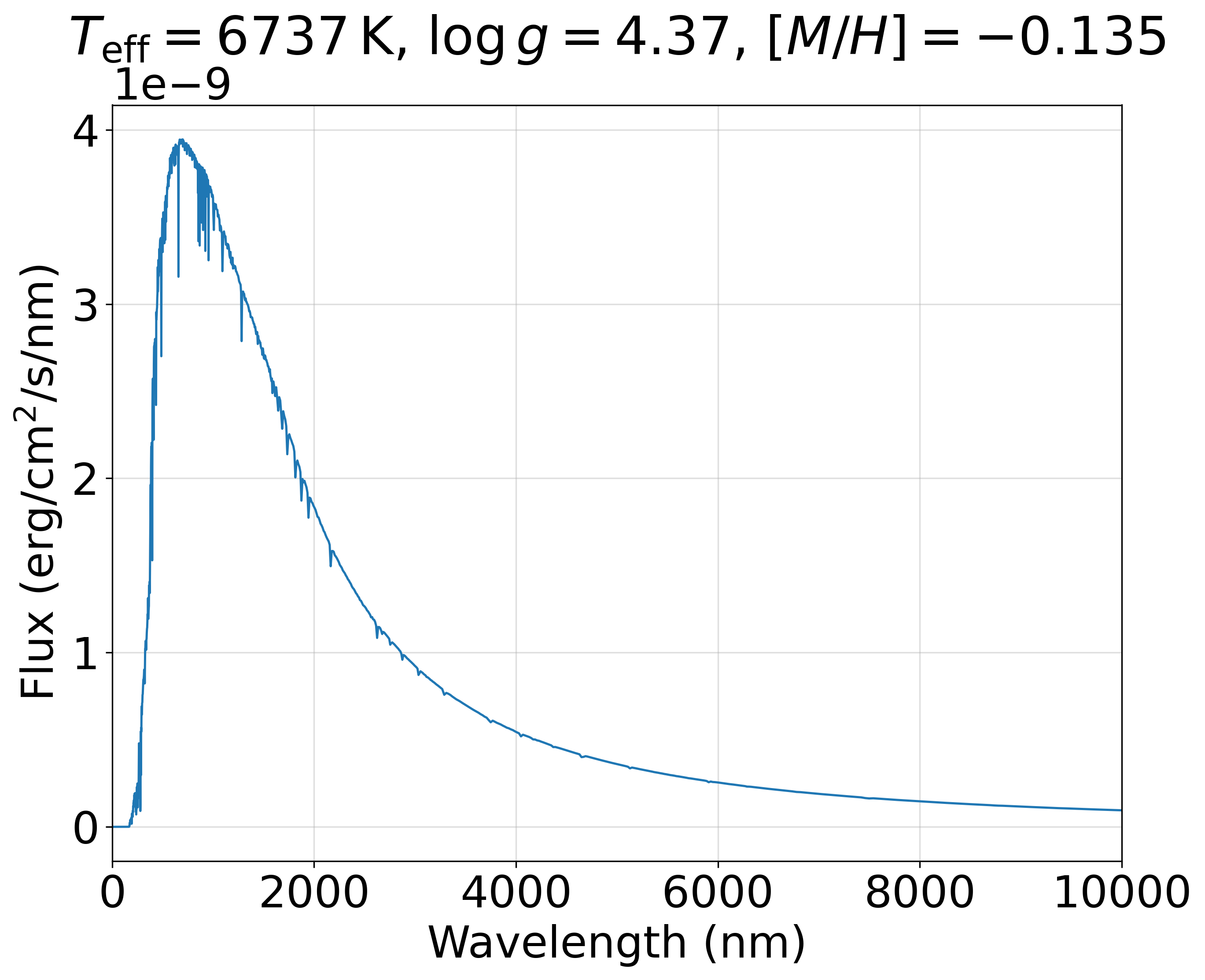} \\
    \includegraphics[width=0.5 \textwidth]{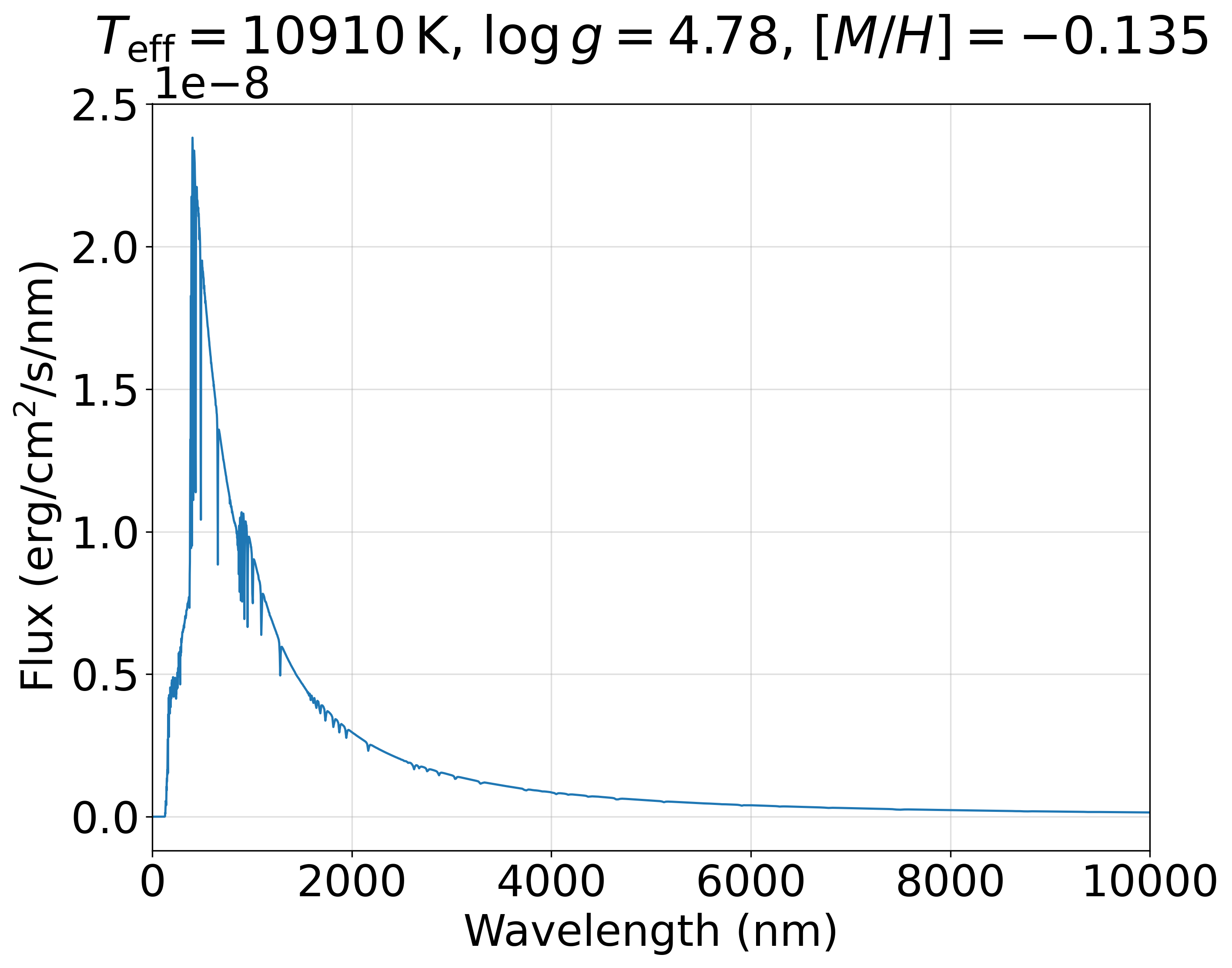} 
    \caption{Disk-integrated spectrum of the convective  (top) and radiative (bottom) nuggets corresponding to the profiles in \fref{f.profile1}.}
    \label{fig:spectra}
\end{figure}

\begin{figure*}
    \centering
    \begin{tabular}{cc}
    \includegraphics[width=0.48 \textwidth]{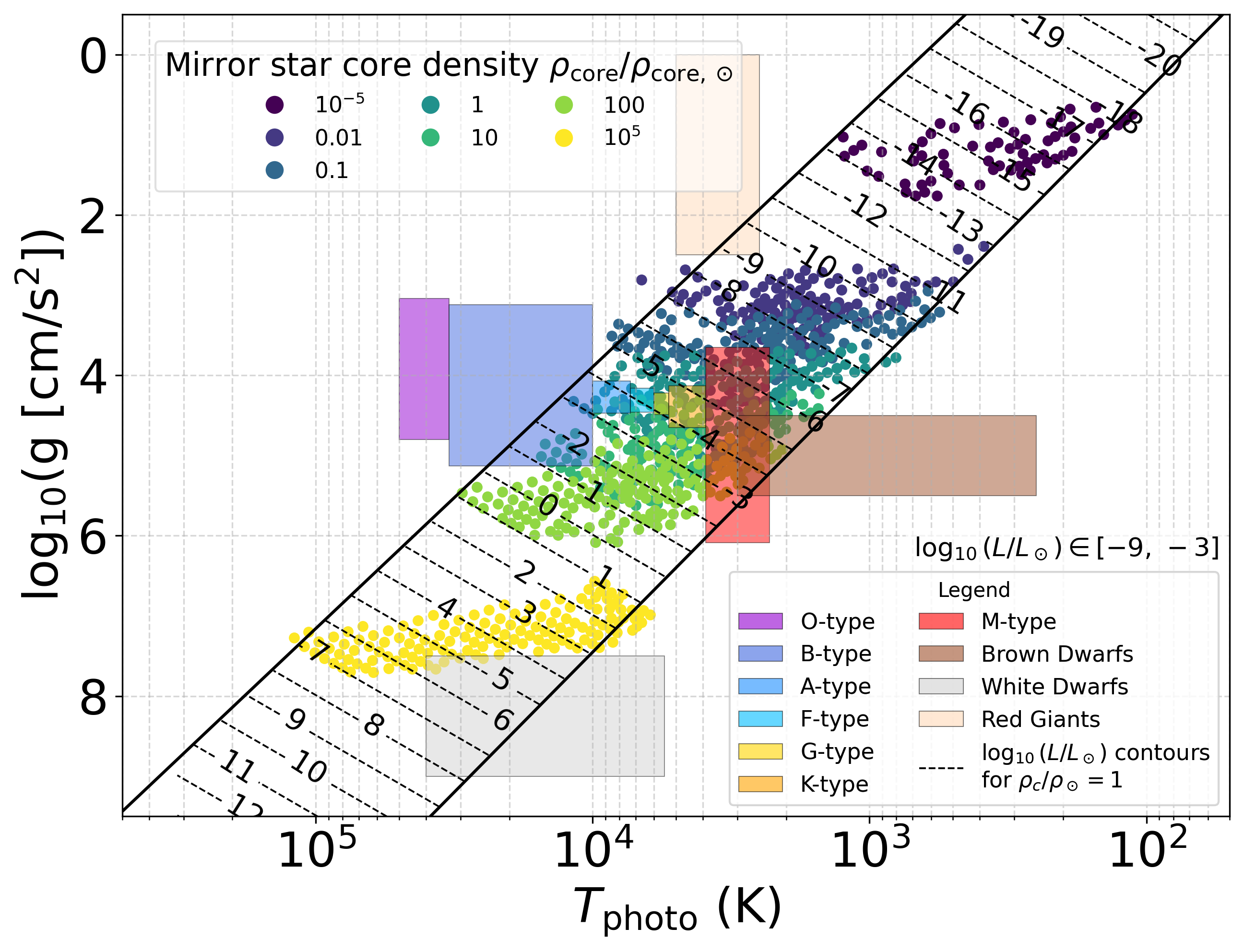}
    &
    \includegraphics[width=0.48 \textwidth]{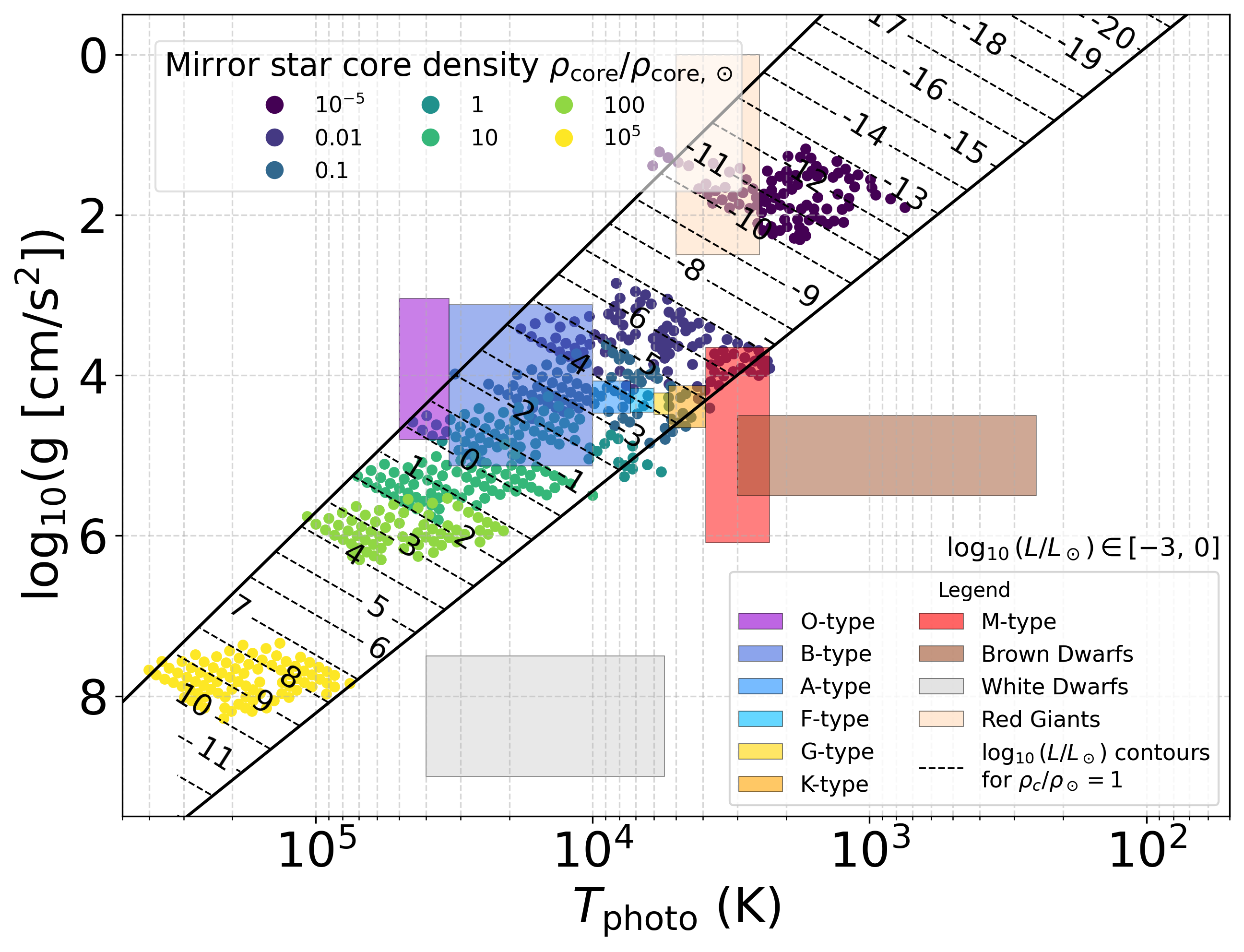}
    \end{tabular}
    \caption{Temperature-surface-gravity diagram for low-luminosity  ($10^{-9} < L/L_\odot < 10^{-3}$) (left) and high-luminosity ($10^{-3} < L/L_\odot < 1$) (right) mirror star nuggets. Our optically thick nugget solutions are shown as circular markers with a color indicating the mirror star core density. Colored rectangles indicate approximate regions populated by different types of Standard Model stars.  
    The solid contours demarcate an approximate signal region for mirror stars in each luminosity range. Dashed contours indicate nugget luminosity $\log_{10} [(L/L_\odot )(\rho_{\mathrm{core}, \odot}/\rho_\mathrm{core})]$.
    }
    \label{fig:Tg}
\end{figure*}

\subsection{Nugget Atmosphere Model}
\label{s.atmosphere}

We modeled the mirror star atmospheres using the MPS-ATLAS (Set 2) stellar atmosphere grid from \cite{Witzke2021}. The MPS-ATLAS (Set 2) grid, based on the stellar models of \cite{Kostogryz2022}, spans a parameter space with metallicities $[\rm M/H] \equiv \log_{10}(M/H)/(M/H)_\odot =  -5.0$ to $1.5$,
effective temperatures $T_\mathrm{eff} = 3500$--$9000$ K, and surface gravities $\log_{10}(g/\mathrm{cm}~\mathrm{s}^{-2}) = 3.0$--$5.0$. 
It contains 34,160 models, each evaluated at 1,221 wavelengths and 24 radial positions across the stellar disc. Set 2 adopts \cite{Asplund2009} abundances and a chemical mixing length that varies with stellar parameters following \cite{Viani2018}.

For each mirror star nugget solution, the surface gravity of the nugget 
was  combined with the computed photospheric temperature $T_\mathrm{photo}$ and the metallicity of interstellar medium $[\rm M/H] = -0.135$ 
to select the appropriate atmosphere model. Disk-integrated spectra were obtained by interpolating the MPS-ATLAS fluxes across wavelength. 
The spectra include prominent absorption lines such as the Balmer series (H$\alpha$, H$\beta$, H$\gamma$, H$\delta$), Ca II H and infrared triplet lines, and the Li I doublet. Figure \ref{fig:spectra} shows sample disk-integrated spectra for two nuggets, one convective and one radiative. 
The spectra are publicly available\textsuperscript{\ref{foot.github}} as 1D arrays of wavelength and disk-integrated flux $F_\nu$ at 1 AU, with units $[\mathrm{erg\,s^{-1}\,cm^{-2}\,Hz^{-1}}]$.

\subsection{Comparison to Regular Stars}

\begin{figure*}[t]
\centering
\begin{minipage}[c]{0.65\textwidth}
  \vspace*{-6mm}
  \centering
  \includegraphics[width=\linewidth]{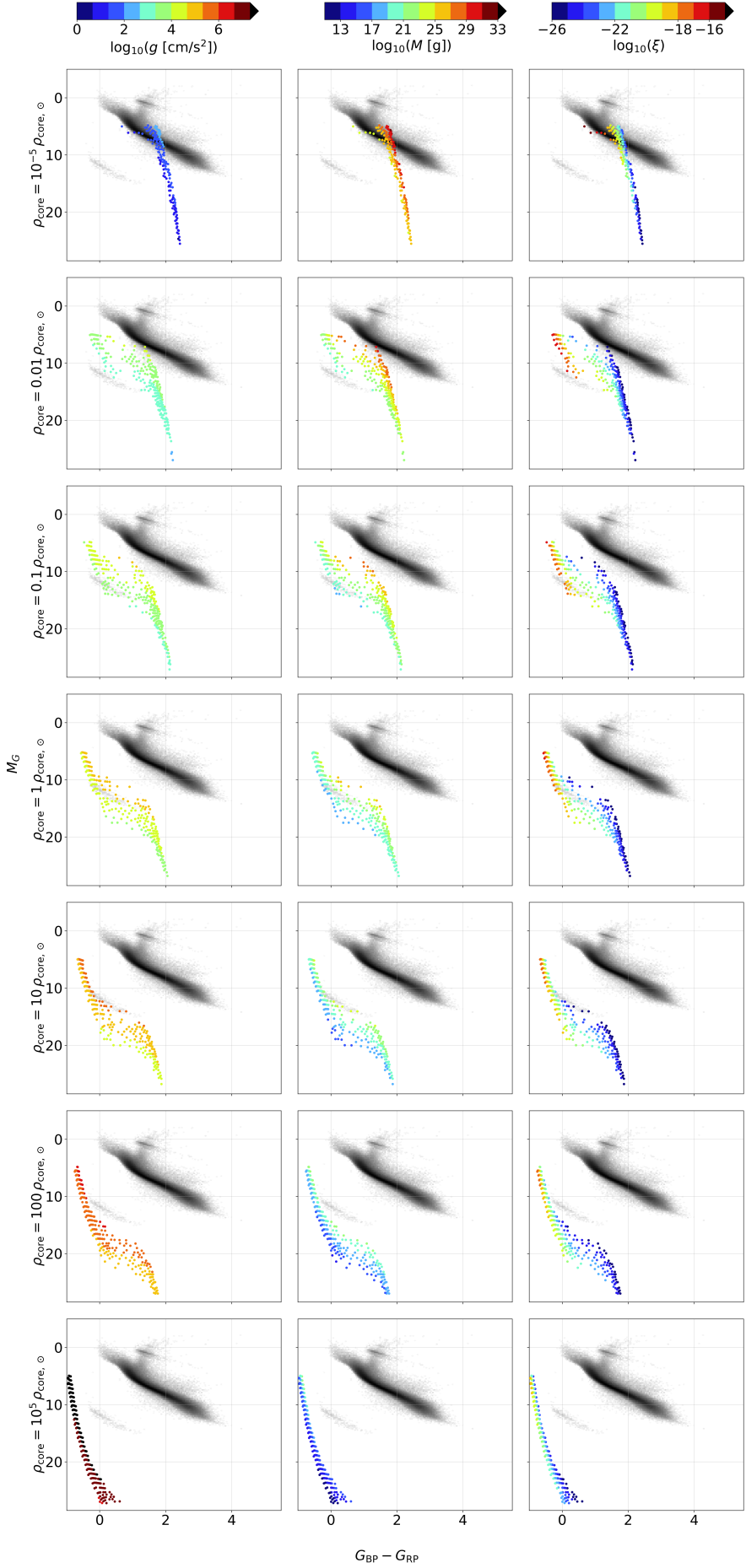}
\end{minipage}\hfill
\begin{minipage}[c]{0.30\textwidth}
  \caption{
        Hertzsprung-Russell diagrams for mirror stars with optically thick nuggets, for different central mirror star densities $\rho_\mathrm{core} / \rho_{\mathrm{core},\odot}$ (different rows). 
        Each panel shows absolute magnitude $M_G$ versus Gaia color index $(G_{\rm BP} - G_{\rm RP})$. 
        Background stars from Gaia are shown in grayscale number density bins, while the mirror star tracks are plotted as colored scatter points. 
        The color of mirror star markers in each scatter point indicates the surface gravity $\log_{10}(g/\mathrm{cm}~\mathrm{s}^{-2})$, nugget mass, and heating rate (left to right columns).}
  \label{fig:stacked_HR}
\end{minipage}
\end{figure*}

\begin{figure*}[t]
\centering
  \includegraphics[width=\textwidth]{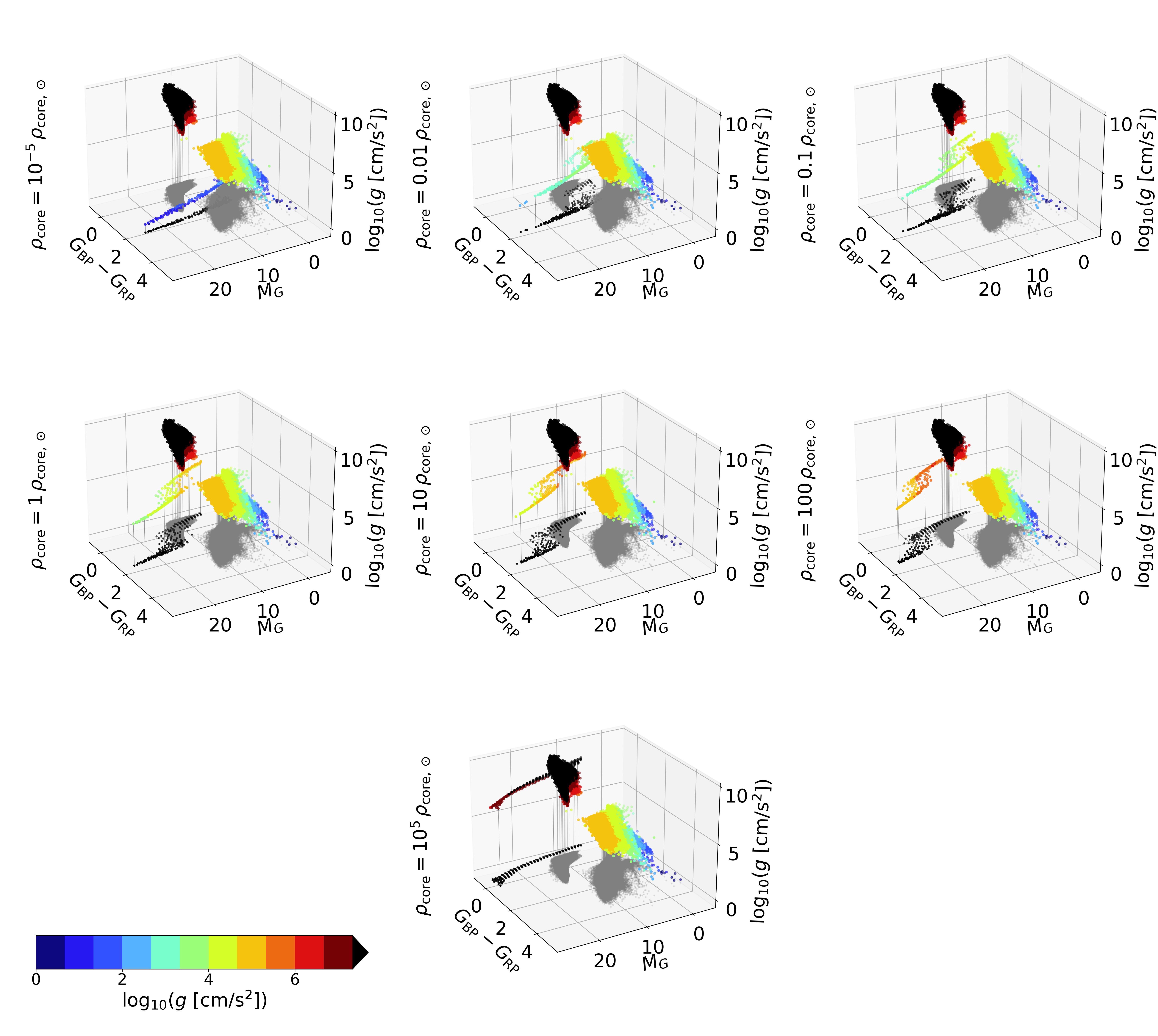}
 \caption{
            3D HR diagrams, showing surface gravity on the vertical axis and in the color of the markers, comparing regular stars (light gray shadows in the HR diagram plane) to optically thick mirror star nuggets (darker shadows) for different $\rho_\mathrm{core}$.  
            The mirror star population generally occupies different regions of this parameter space than regular stars.
            }
  \label{fig:HR3D}
\end{figure*}

There are a few main ways in which optically thick nuggets can be distinguished from regular stars. Most importantly, the nuggets will generally occupy a different region of the $\{T, \mathcal{L}, \log g\}$ parameter space than regular stars. The first two parameters, of course, just define an Hertzsprung-Russell (HR) diagram, while $\log g$ can be ascertained from  the broadening of absorption lines in the spectrum. 

Another distinguishing factor is that unlike stars that form from a collapsing rotating gas cloud, mirror star nuggets that accumulate from ISM capture are likely to have no or much lower rotation rates, which can be discerned from periodic photometric variability and from the Doppler broadening of their absorption lines. 
Furthermore, the material within a mirror star nugget will not have experienced nuclear reactions since its accumulation from the interstellar medium, and therefore it will not be depleted of the light elements lithium, beryllium, and boron by proton capture.  Since many of the nearby objects in the HR diagram display surface depletion of these elements \citep{Boesgaard_2023}, this may provide a useful discriminant.

Finally, in both optically thick and thin nuggets, dark photons from the mirror star core would be converted directly to X-rays via elastic Compton scattering of Standard Model electrons in the nugget, leading to an X-ray signal with a frequency that is characteristic of the mirror star core temperature.

In this paper we obtained detailed predictions for the first of these possible signals and will hence focus on them in our discussion, but the others will likely also be helpful for actual mirror star searches.

\fref{fig:Tg} shows where mirror star nuggets fall in the surface temperature - surface gravity plane, compared with the regions broadly associated with different star types. Clearly mirror stars can be discernible from regular stars even in this 2D projection of the full $\{T, \mathcal{L}, \log g\}$ parameter space. The top (bottom) plot shows low-luminosity (high-luminosity) nuggets with $10^{-9} < L/L_\odot < 10^{-3}$ ($10^{-3} < L/L_\odot < 1$), and different color markers distinguishing different mirror star core densities $\rho_\mathrm{core}/\rho_\mathrm{core,\odot}$. 
There are clearly defined signal regions for low- and high-luminosity nuggets, shown as the diagonal band marked with black boundaries. Dashed contours inside the band show the nugget luminosity, assuming a central density equal to the standard solar one. For different mirror star central densities, multiply that luminosity by $\rho_\mathrm{core}/\rho_\mathrm{core, \odot}$.

HR diagrams comparing optically thick mirror star nuggets to regular stars from the Gaia catalogue DR3 are shown in \fref{fig:stacked_HR}. 
For the nuggets, the absolute Gaia $G$-band magnitude $M_G$ was computed as
$M_G = 4.83 - 2.5 \log_{10} (L/L_\odot)$,
where $L$ is the bolometric luminosity of the mirror star. 
The Gaia color index $G_{\rm BP}-G_{\rm RP}$ was estimated by inverting the following relation from \cite{Jordi2010}:
\begin{equation}
\log_{10}(T_{\rm eff}) = 3.999 - 0.654\,C + 0.709\,C^2 - 0.316\,C^3,
\end{equation}
with $C = G_{\rm BP}-G_{\rm RP}$, providing a mapping between effective temperature and observed color. High-mass nuggets can be degenerate with stars in this plane, emphasizing the importance of stellar spectra to determine surface gravity and other features like rotation rates and light elemental abundances. However, lower-mass nuggets are clearly distinct from main sequence and white dwarf populations. Note that the mirror star signal region in the HR diagram is much larger than the signal region assumed in the earlier Gaia mirror star search by \cite{Howe:2021neq}, which used a very basic approximation for the expected emission spectrum of optically thick nuggets. This can also be compared with the distribution of optically thin mirror star nuggets in the HR diagram \citep{Armstrong:2023cis}, which are confined to the well-defined color band $G_{BP} - G_{RP} \in (0.4, 1.4)$, mostly just below the white dwarf population.

In \fref{fig:HR3D} we attempt to show the position of mirror stars in the full 3D $\{T, \mathcal{L}, \log g\}$  parameter space. This illustrates that with absolute magnitude and spectral measurements, almost all mirror stars can likely be distinguished from standard stars.
Furthermore, both Figures~\ref{fig:stacked_HR} and~\ref{fig:HR3D} make clear that combining astrometric and spectral measurements of a mirror star candidate would in principle allow for  $\{\rhocms, \xi, \Mnugget\}$ to be uniquely determined. (This can be easily seen from \fref{fig:stacked_HR}, where for a given location on the HR diagram, the surface gravity measurement breaks the degeneracy between different possible mirror star core densities.) 
A single mirror star observation would therefore not only constitute a direct discovery of dark matter, but also a supply information on the dark sector microphysics (the heating rate) and mirror stellar physics (the mirror star core density).

\section{Conclusion}
\label{s.conclusion}

Mirror stars are a spectacular but surprisingly general prediction of a dissipative dark matter subcomponent.
Even in the absence of non-gravitational interactions with the SM, such exotic compact objects are promising targets for gravitational wave and microlensing searches~\citep{Winch:2020cju, Perkins:2025hfr, Hippert:2021fch, Hippert:2022snq, Pollack:2014rja, Shandera:2018xkn, Singh:2020wiq, Gurian:2022nbx, Fernandez:2022zmc}. 
However, similarly general theoretical arguments based on symmetries and naive dimensional analysis suggest the existence of a small kinetic mixing between the dark photon and our photon. This results in mirror stars capturing SM matter from the ISM. The resulting `nugget'  accumulates in their cores, gets heated up, and emits detectable electromagnetic radiation.
Mirror stars are therefore attractive targets for optical/IR and X-ray ray telescope searches, which offer the possibility of directly discovering exotic compact dark matter objects, or of placing strict limits on their abundance  within the Milky Way. 

In this paper, we analyzed the thermal emissions of mirror stars that accumulate enough ISM material for the captured nugget to be optically thick. This complements a previous analysis for optically thin nuggets \citep{Armstrong:2023cis}. The result, made publicly available,\textsuperscript{\ref{foot.github}}  is a library of mirror star optical/IR emission spectra that span the effective signature parameter space  of nugget mass, mirror star central density, and heating rate
$(M_\mathrm{nugget}, \rho_\mathrm{core}, \xi)$, see \fref{f.parameterspace}.
This can be used directly to supply templates for future telescope searches. 

Equally important is a general understanding of what distinguishes mirror star electromagnetic emissions from standard stars and other astrophysical objects. 
Optically thick nuggets appear naively star-like, but generally occupy very distinct regions in the surface temperature, luminosity and surface gravity parameter space
$(T, L, \log g)$, see \fref{fig:HR3D}. These differences are also apparent when projecting down to 2D HR diagrams, see \fref{fig:stacked_HR}, or the temperature-surface gravity plane, see \fref{fig:Tg}. 
Additional features, such as the expected lack of rotation, and the presence of lighter elements in nuggets that would be depleted in regular stars, can provide further checks on the mirror star nugget nature of a given observation.
Optically thin nuggets, as determined by \cite{Armstrong:2023cis}, are similarly detectable. In an HR diagram, they occupy a well-defined color band below the white dwarf population. Spectral analysis will reveal the continuum nature of their emission, as distinct from the more blackbody-like emission of stars, but the ratio of various emission lines will reveal their much higher density compared to standard nebulae, again providing an orthogonal handle for distinguishing them from standard astrophysical objects.

A combination of astrometrical and spectral analyses should therefore reveal any such anomalous objects in our stellar catalogues. A confirmed observation could not only constitute a dark matter discovery, but also supply detailed information about the dark sector micro- and stellar physics by, in principle, uniquely determining $\{\rhocms, \xi, \Mnugget\}$.
Conducting such a search is now possible, and will be an important future study. 

Other important future directions include a similarly general analysis of the nugget's X-ray emission via direct Compton conversion of thermal dark photons in the mirror star core, 
which would not only constitute an additional discovery handle but also supply direct information about the mirror star core temperature. 
An exploration of mirror star stellar physics for various concrete dissipative dark matter models is also highly motivated, as it would allow us to associate  specific regions of our effective mirror star signature parameter space with different dark matter scenarios. 

Mirror stars represent a unique opportunity to unambiguously discover dark matter directly with telescopes. Any potential candidate that displays the distinctive spectral features we identify, alongside the characteristic X-ray emissions from dark photon conversion, would not only confirm the existence of dissipative dark matter but would also supply information about its detailed properties. This exciting potential for discovering and characterizing new physics beyond the Standard Model, and shedding light on one of the most important mysteries of particle physics and cosmology, should serve as strong motivation to search for mirror stars with every means at our disposal.

\section*{Acknowledgments}
We thank Philip Mocz for detailed discussions on the MESA stellar evolution code.
The work of FC, SW and DC was in part supported by Discovery Grants from the Natural Sciences
and Engineering Research Council of Canada, the
Canada Research Chair program, 
the Ontario Early Researcher Award, and the University of Toronto McLean Award. 
The work of CM was supported by an NSERC Discovery Grant.

\bibliography{ref}

\end{document}